# A Dynamic Web Page Prediction Model Based on Access Patterns to Offer Better User Latency

Debajyoti Mukhopadhyay[1, 2]    Priyanka Mishra[1]    Dwaipayan Saha[1]    Young-Chon Kim[2]

[1] Web Intelligence & Distributed Computing Research Lab, Techno India (West Bengal University of Technology)
EM 4/1, Salt Lake Sector V, Calcutta 700091, India
Emails: {debajyoti.mukhopadhyay, dwaipayansaha, priyanka147}@gmail.com

[2] Chonbuk National University, Division of Electronics & Information Engineering
561-756 Jeonju, Republic of Korea;  Email: yckim@chonbuk.ac.kr

*ABSTRACT*
*The growth of the World Wide Web has emphasized the need for improvement in user latency. One of the techniques that are used for improving user latency is Caching and another is Web Prefetching. Approaches that bank solely on caching offer limited performance improvement because it is difficult for caching to handle the large number of increasingly diverse files. Studies have been conducted on prefetching models based on decision trees, Markov chains, and path analysis. However, the increased uses of dynamic pages, frequent changes in site structure and user access patterns have limited the efficacy of these static techniques. In this paper, we have proposed a methodology to cluster related pages into different categories based on the access patterns. Additionally we use page ranking to build up our prediction model at the initial stages when users haven't already started sending requests. This way we have tried to overcome the problems of maintaining huge databases which is needed in case of log based techniques.*

*Keywords*
Levels, Classes, Product Value, Prediction Window, Date of modification, Page rank, links, prediction model, Predictor, Update Engine

## 1. INTRODUCTION

The exponential proliferation of Web usage has dramatically increased the volume of Internet traffic and has caused serious performance degradation in terms of user latency and bandwidth on the Internet. The use of the World Wide Web has become indispensable in everybody's life which has also made it critical to look for ways to accommodate increasing numbers of users while preventing excessive delays and congestion. Studies have been conducted on prefetching models based on decision trees, Markov chains, and path analysis. [1][2][4] There are several factors that contribute to the Web access latencies such as:

- Server configuration
- Server load
- Client configuration
- Document to be transferred
- Network characteristics

Web Caching is a technique that made efforts to solve the problem of these access latencies. Specially, global caching methods that straddle across users work quite well. However, the increasing trend of generating dynamic pages in response to HTTP requests from users has rendered them quite ineffective. The following can be seen as the major reasons for the increased use of dynamic Web pages:

1. For user customized Web pages the content of which depends on the users' interests. Such personalized pages allow the user to reach the information they want in much lesser time.

2. For pages that need frequent updating it is irrational to make those changes on the static Web pages. Maintaining a database and generating the content of the Web pages from the database is a much cheaper alternative. Pages displaying sports updates, stock updates weather information etc. which involve a lot of variables are generated dynamically.

3. Pages that need a user authentication before displaying their content are also generated dynamically, as separate pages are generated as per the user information for each user.

then sending those predicted requests to the user before he/she actually makes the request.

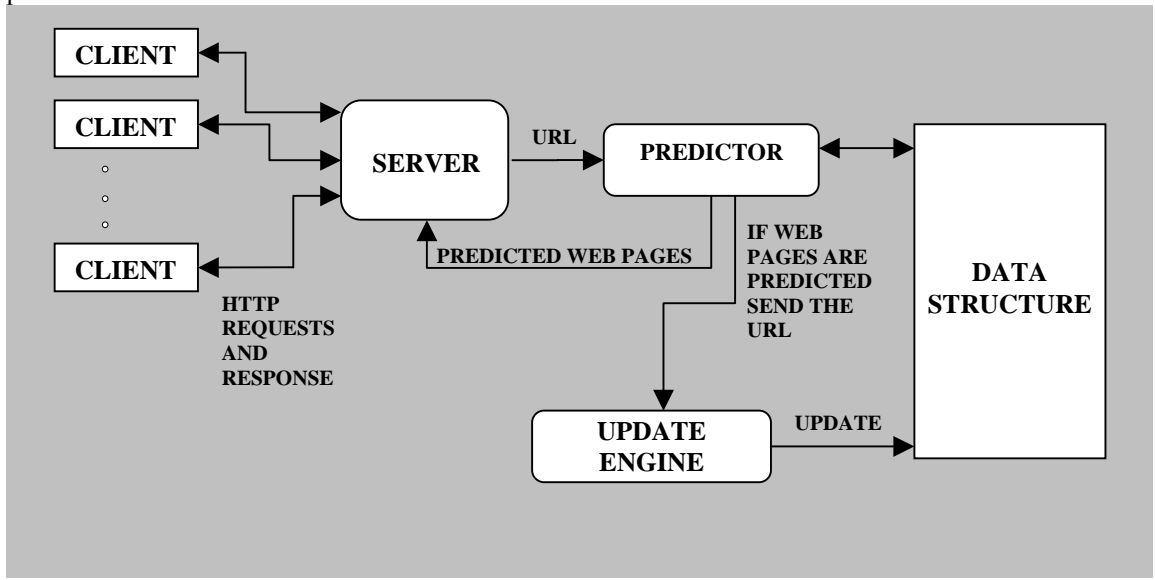

**Figure 1:** Architecture of the Prediction Model

This trend is increasing rapidly.
4. All response pages on a secure connection are generated dynamically as per the password and other security features such as encryption keys. These pages expire immediately by resetting the Expire field and/or by the Pragma directive of 'nocache' in the HTTP header of the server response, to prevent them from being misused in a Replay attack.

As the Internet grows and becomes a primary means of communication in business as well as the day to day life, the majority of Web pages will tend to be dynamic. In such a situation traditional caching methods will be rendered obsolete. The dynamic pages need a substantial amount of processing on the server side, after receiving the request from the client and hence contribute to the increase in the access latency further.

An important prefetching task is to build an effective prediction model and data structure for predicting the future requests of the user and

The organization of rest of the paper is as follows: our methodology is presented in Section 2, in Section 3 the Experimental Setup is described, Section 4 shows the Experimental Results, Section 5 has the concluding remarks. There is a list of references included at the end of this document.

## 2. PROPOSED METHODOLOGY
### 2.1 Prediction Model
Links are made by Web designers based on relevance of content and certain interests of their own. In our model, we classify Web pages based on hyperlink relations and the site structure. We use this concept to build a category based dynamic prediction model. For example in a general portal www.abc.com all pages under the movies section fall under a single unique class. We assume that a user will preferably visit the next page, which belongs to the same class as that of the current page. To apply this concept we consider a set of dominant links that point to pages that define a particular category. All the pages followed by that particular link remain in the same class. The pages are categorized further

into levels according to the page rank in the initial period and later, the users' access frequency.[6][7]

The major problem in this field is that, the prediction models have been dependent on history data or logs.[5] They were unable to make predictions in the initial stages.[3] We present the architecture of our model in Figure 1. Our model is not dependent on log data rather it is built up using ranking of pages and updated dynamically as HTTP requests from the users arrive.[6][8][9][10][11]

HTTP requests arrive at the Predictor. The Predictor uses the data from the data-structure for predicting, and after predicting the forthcoming requests, passes the requested URL to the Update Engine to update the data structure. For constructing the initial model we define a subset of the set of total pages in the site as dominant pages. Based on these dominant pages classification of the pages in the site is done. For example in a general portal www.abc.com sports.html may be a dominant page which is the base page for the 'sports' class. The candidates for dominant pages may be choosen manually by the site administrator or all the links in the home page may be considered as dominant pages when the server is started. The algorithm to create the initial model is as shown below:

```
Input: The set of URLs U= {u1, u2, u3…
uk}, set of dominant pages={d1,d2,…,dn}.
Output: The prediction model T.

An empty array called common-page is used
for holding pages which are linked to
more than one class.
stack1, stack2 are empty stacks.

1: Based on the page ranks, assign level
   numbers to pages.
2: Put all the dominant pages in
   stack1, and assign them a unique class
   number.
3: while (stack1 is not empty),
   pop the first element from stack1,name
   it P.
   for(all pages pointed by P)
   if(any page is assigned a class
   number) then
   if(class number is same as P) then
   do nothing
   else
   add that page to common-page.
   end else
   else
   a. assign the class number of P as the
   class number of that page
   b. push that page in stack2.
   end else.
   end for
   pop all elements from stack2 and push
   them in stack1 in reverse order.
   end while.
4: for(each page in common-page) reassign
   the class number same as that of the
   class having  maximum number of links
   pointing to it.
```

In our model shown in Figure 2 we categorize the users on the basis of the related pages they access. Our model is divided into levels based on the popularity of the pages. Each level is a collection of disjoint classes and each class contains related pages. Each page placed in higher levels has higher probability of being predicted. Mathematically the model may be represented as:

```
Let T is the Prediction Model,
Let C = {C₁, C₂, C₃,…,Cₙ  } is the set of
planes, where n= Number of planes.
For every element Cᵢ in C there exists,
CUᵢ = {U₁, U₂, U₃ ,…,Uₘ }which is a set of
URLs in plane Cᵢ

And i = 1,2,…,n. and k≠ j for all Uₖ, Uⱼ
belonging to C where k, j = 1,2,…,m

Also,
```

$$\bigcap_{P=1}^{P=n} C_p = \{ \ \}$$

$$\bigcup_{P=1}^{P=n} C_p = T$$

Each $C_i$ in C has its own level number.

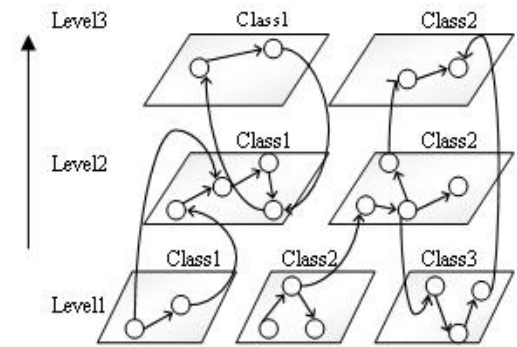

**Figure 2:** Representation of the Model

The disjoint classes signify the categorization of the pages accessed by the users. Each level signifies the possibility of the page to be accessed in the future. Higher the level higher is the possibility. The pages in the various classes are promoted to the higher levels based on the number of accesses to that page by the user. The next request for a page is predicted according to its presence in a higher level than the current page that points to it. More than one page is predicted and sent to the user's cache depending upon the presence of links in the higher levels.

After calculating page ranks, a normalized value in the range of 1 to p is assigned to each page where p is the number of pages in the site. For storage reasons the number of levels is restricted to a predefined constant value L, where typically $L=\lceil\sqrt{p}\rceil$. We further divide the p pages into L sets. For each set, classes are formed depending upon the actual links present between them. Thus pages are categorized into disjoint classes "C." Each level and class is assigned a distinct number. In order to search for the presence of a page, the URL name is used as a key to the hash table data- structure.

Since we're working with a range of values for a level, we assign a counter to all the pages except those already in the uppermost level. For each request the counter is incremented when it reaches L, the page is promoted to the next higher level. Pages may traverse between levels when any of the following conditions occur:
1. The page is demoted to a lower level when the time stamp value assigned to it expires.
2. The page is promoted to a higher level if it has been modified recently.

This is discussed further in Sub-section 2.3.

**2.2 Predictor**
All required information about the pages of the Website is indexed using their URLs in a Hash table where a URL acts as the key. When a request is received a search on the hash table is conducted and the information thus obtained is analyzed in the following manner:
```
1. Get the level and class number of
   the requested URL.
2. Get the links associated with the
   page and also fetch their
   respective level and class
   numbers.
3. Determine Prediction-Value(P-
   value) pairs for the entire
   candidate URLs, where a P-value
   pair is defined as [Level, Rank].
4. Sort the links of the requested
   URL according to the following
   precedence relations defined by
   their P-Value pairs:

There can be four types of
precedence relations between two
P-value pairs (Li,Ri) and
(Lj,Rj):
```
i) (Li,Ri) <· (Lj,Rj) when
    a. Li<Lj & Ri<Rj
    b. Li=Lj & Ri<Rj
    c. Li<Lj & Ri=Rj
    d. Li<Lj & Ri>Rj

implies (Lj,Rj) precedes (Li,Ri).

ii) (Li,Ri) ·> (Lj,Rj) when
    a. Li>Lj & Ri<Rj
    b. Li>Lj & Ri=Rj
    c. Li>Lj & Ri>Rj
    d. Li=Lj & Ri>Rj

implies (Li,Ri) precedes (Lj, Rj).

iii) (Li, Ri) || (Lj,Rj) when:
    a. Li>Lj & Ri<Rj
    b. Li<Lj & Ri>Rj

implies (Li,Ri) and (Lj,Rj) are incomparable.

iv) (Li,Ri) ≈ (Lj,Rj) when:
    a. L1=L2 & R1=R2

implies (Li,Ri) & (Lj,Rj) are equivalent.

```
5. Compare the links' level number
   with the URLs' level number.
6. Compare the class numbers of the
   links with that of the requested
   URL. The link having the same
   class number will get preference.
7. The links in the higher levels
   are the predicted links to be
   sent to the users' cache.
```

**2.3 Update Engine**
In the updating process we adjust the counter value and decide whether the page should go to a higher level, class numbers are assigned at the

initial stage and remain static. The process may be described as follows:

```
1. Check     the     local     counter
   associated  with  the  requested
   URL.
2. If the counter value is less than
   (L-1) then increment the counter
   Else
   Fetch the current level number of
   the URL. Let it be L.
   Increment L.
3. Reset the counter and time stamp.
```

The Update Engine also checks periodically for a page that is present in a higher level and has not been accessed for a long duration to relegate it to a lower level according to a predetermined threshold value. This periodic process compares the timestamp of all the pages with this threshold value and demotes those pages which exceed this threshold. There's another periodic check that checks the last date of modification of the page. If there is recent modification then the page is raised to a higher level. This is done as a recently modified page always has higher probability of being accessed by the user.

## 3. EXPERIMENTAL SETUP
The prediction model is implemented using a link data-structure which is shown in Figure 3:

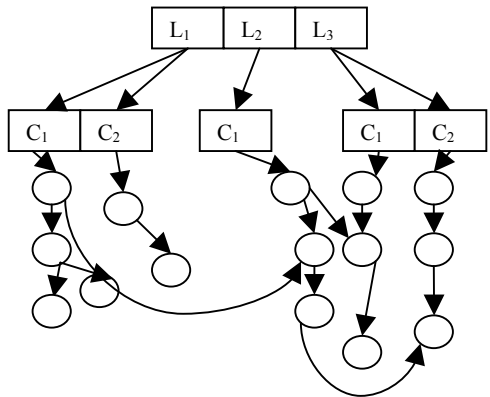

**Figure3:** Data-Structure Representing the Prediction Model

This data-structure represents the categorization of the URLs where the levels $L_1, L_2,..L_n$ act as index of the respective classes $C_1, C_2....C_n$. Each $L_i$ where $i = 1,…, n$ is the root for its classes and each class is the root for the respective URL trees.

This data-structure is implemented in the form of a hash table with URLs being used as the key. Table 1 shows the implementation of the above data-structure

| Key | URL | LC | L# | C# | TS | DM | Links |
|-----|-----|----|----|----|----|----|-------|
| A1  | A   | 2  | 1  | 2  | Xx | Yy | A4,.. |
| A2  | B   | 0  | 1  | 2  | Xx | Yy | A8,.. |
| A3  | C   | 2  | 1  | 3  | Xx | Yy | A1,.. |
| A4  | D   | 1  | 2  | 3  | Xx | Yy | A5,.. |
| A5  | E   | 3  | 2  | 4  | Xx | Yy | A7,.. |
| A6  | F   | 1  | 2  | 4  | Xx | Yy | A5,...|

**Table 1:** Tabular Representation of the Data-Structure

Following are the brief description of each of the labels used in the data-structure:
- 'Key' represents the key of the current row in the hash table.
- 'URL' represents the Web address of the page.
- 'LC' represents the local counter associated with the page which represents the number accesses made to the page in a particular level. When this value reaches (L–1) the page is promoted to a higher level and this counter is reset.
- 'L#' is the level number and 'C#' is the class number.
- 'TS' is the timestamp associated with the URL that represents the duration for which the page has been in a particular level.
- 'DM' represents the last date of modification of the page.
- 'Links' represents the list of links to which the current URL points.

When a request is received the row for the requested URL is fetched from the hash table using URL as the key. The level and class numbers are obtained. The links are fetched from the hash table and their class and level numbers are also fetched. Thus we can make predictions on the basis of level, rank and class values of the linked URLs.

## 4. EXPERIMENTAL RESULTS

We have used about one hundred Web-pages residing in our Web-Server for generating the test results. We examined the hit percentage vs. user session as per the prediction window size. The size of the prediction window was taken as two and three considering the number of pages in our test environment. Size of a prediction window indicates the number of Web-Pages sent to the Client-cache by the Web-Server while predicting the pages. The hit percentage remained consistent throughout the testing period including the initial stages.

The average hit percentage was found to be around 35% with a prediction window size of 2 and 51% with a prediction window size of 3, an improvement of around more than 15%. The following chart is plotted with sessions recorded at different intervals during the testing period with variable prediction window sizes (Chart 1).

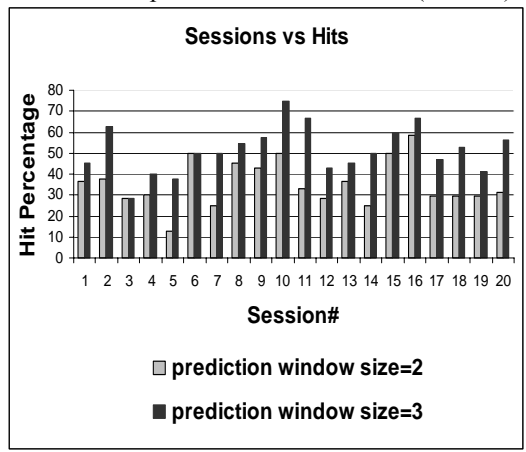

**Chart 1:** Chart showing comparative hit ratio with different prediction window sizes.

## 5. CONCLUSION & FUTURE WORK

In most of the cases prediction of Web pages is done using logs and history data which require a huge amount memory to implement. Another problem that is seen is the inability to build up the prediction model in the initial stages when no log or history data is available. The use of page ranking in our model enables us to build up our prediction model in the initial stages and make predictions right away. Henceforth our model updates itself as per the access patterns of users.

Categorizing the users into different classes also helps as we don't have to keep track of each user as all access patterns are maintained in the form of sessions. Updating the model dynamically according to access patterns of users as well as changes in the content of the Website is computationally cheaper as it doesn't put extra load on the Web traffic for requesting or maintaining extra information. In our next step we would like to build Markov and Decision tree models from the knowledge discovered through access patterns and study their performance.

## REFERENCES

1. Xin Chen, Xiaodong Zhang; *Popularity-Based Prediction Model for Web Prefetching*; IEEE Computer;vol.36;no.3;2003.www.cs.wm.edu/hpcs/WWW/HTML/publications/papers/TR-03-1.pdf.
2. Themistoklis Palpanas; *Web Prediction using Partial Match Prediction*; Department of Computer Science; University of Toronto; Technical Report CSRG-376. www.it.iitb.ac.in/~it620/papers/WebReqPatterns.pdf.
3. Brian D. Davison; *Learning Web Request Patterns*; Department of Computer Science and Engineering; LehighUniversity;www.cse.lehigh.edu/~brian/pubs/2004/Webdynamics/chapter.pdf.
4. *Dario Bonino, Fulvio Corno; Giovanni Squillero; Politecnico di Torino;* **An Evolutionary Approach to Web Request Prediction**; *The Twelfth International World Wide Web Conference 20-24 May 2003;Budapest,Hungary.ht*tp://www.cad.polito.it/pap/db/cec03a.pdf.
5. Z. Su, Q. Yang, Y. Lu, H. Zhang; *WhatNext: A Prediction System for Web Requests using N-gram Sequence Models*; Proc. of the First International Conference on Web Information System and Engineering Conference;2000; pp.200-207.
6. Lawrence Page, Sergey Brin, Rajeev Motwani, Terry Winogra; *The PageRank Citation Ranking: Bringing Order to the Web*; *http://google.stanford.edu/~*backrub*/pageranksub.ps*
7. J. Kleinberg; *Authoritative sources in a hyperlinked environment*; In Proc. Ninth Ann. ACM-SIAM Symp. Discrete Algorithms, ACM Press, 1998; pp. 668-677.
8. Debajyoti Mukhopadhyay, Debasis Giri, Sanasam Ranbir Singh; *An Approach to Confidence Based Page Ranking for User Oriented Web Search*; SIGMOD Record, Vol.32, No.2, June 2003; pp. 28-33.
9. Debajyoti Mukhopadhyay, Sanasam Ranbir Singh; *An Algorithm for Automatic Web-Page Clustering using Link Structures*; IEEE INDICON 2004 Proceedings; IIT Kharagpur, India; 20-22 December 2004; pp. 472-477.
10. Debajyoti Mukhopadhyay, Pradipta Biswas; *FlexiRank: An Algorithm Offering Flexibility and Accuracy for Ranking the Web Pages*; ICDCIT 2005; India; LNCS, Springer-Verlag, Germany; 22-24 December 2005; pp. 308-313.
11. Sergey Brin and Lawrence Page; *The anatomy of a large-scale hypertextual Web search engine*; *In* Proeedings. of WWW Conf.*, 1998*.